\documentclass[11pt,a4paper]{article}
\pdfoutput=1 % if your are submitting a pdflatex (i.e. if you have
\usepackage{jheppub}

\usepackage[utf8]{inputenc}
\usepackage[T1]{fontenc}
\usepackage{lmodern}
\usepackage{fontawesome}
\usepackage{hyperref}
\usepackage{tikz}
\usepackage{subcaption}
\usepackage{mathtools}
\usepackage{xcolor}
\usepackage{amsmath,amssymb}
\preprint{ \vbox{\hbox{IPARCOS-UCM-24-022}}}
\usepackage{tikz-feynman} 

\usetikzlibrary{positioning}
\usetikzlibrary{automata,positioning,arrows,matrix,backgrounds,calc}
\usetikzlibrary{decorations.text}
\usetikzlibrary{decorations.pathmorphing}
\usetikzlibrary{shapes.geometric}
\usetikzlibrary{arrows.meta}

\usepackage{pgf}
\usepackage{relsize}

\usetikzlibrary{decorations,patterns,arrows.meta}

\usetikzlibrary{decorations.pathreplacing,decorations.markings}

\tikzset{
    ->-/.style={decoration={
  markings,
  mark=at position .65 with {\arrow{Latex}}},postaction={decorate}},
    -<-/.style={decoration={
  markings,
  mark=at position .6 with {\arrow{<}}},postaction={decorate}},
    ->/.style={decoration={
  markings,
  mark=at position .4 with {\arrow{>}}},postaction={decorate}},
}
\title{Di-Higgs production via Axion-Like Particles}

\author[a]{Fabian Esser,}
\author[b]{Maeve Madigan,}
\author[a,d]{Alexandre Salas-Bern\'ardez,}
\author[a]{Veronica Sanz}
\author[c]{and Maria Ubiali}

\affiliation[a]{Instituto de F\'isica Corpuscular (IFIC), Universidad de Valencia-CSIC, E-46980 Valencia, Spain}
\affiliation[b]{Institut für Theoretische Physik, Universität Heidelberg, Germany}
\affiliation[c]{DAMTP, University of Cambridge, Wilberforce Road, Cambridge CB3 0WA, UK }
\affiliation[d]{Dept.  Análisis Matemático y Matemática Aplicada and IPARCOS, Univ. Complutense de Madrid, Plaza de las Ciencias 3, 28040 Madrid, Spain}

\emailAdd{esser@ific.uv.es}
\emailAdd{madigan@thphys.uni-heidelberg.de}
\emailAdd{alexsala@ucm.es}
\emailAdd{veronica.sanz@uv.es}
\emailAdd{m.ubiali@damtp.cam.ac.uk}

\newcommand{\hepdata}{{\tt HEPData}\,\,}
\newcommand{\madgraph}{{\tt MadGraph5\_aMC@NLO}\,\,}

\abstract{Due to the pseudo-scalar nature of the axion-like particle (ALP), the CP-conserving production of two Higgs bosons via the ALP necessarily involves an additional $Z$ or $\gamma$ boson. We examine the existing constraints from di-Higgs searches at Run 2 of the LHC and find that, despite the presence of extra objects in the final state, these searches are sensitive to a combination of ALP couplings to gluons and three-bosons in the TeV scale range.  
Additionally, we propose a specialized search strategy incorporating an energetic leptonic $Z$ boson. This refined ALP-induced production process would allow for the identification of the $h\, h \, \to $ 4 $b$-jet final state and could potentially probe the TeV scale using data from Run 2 of the LHC. 
This production process can also occur through a coupling between the top quark and the ALP. We translate the current constraints on di-Higgs production into new limits on the ALP-top coupling. }
\begin{document}

\maketitle

\section{Introduction}
\label{sec:intro}
The Standard Model (SM) of particle physics has enjoyed remarkable success in describing the fundamental constituents of matter and their interactions. However, it is widely recognized that this framework remains incomplete, with numerous phenomena such as dark matter, dark energy, and the matter-antimatter asymmetry of the universe remaining unexplained. Among the proposed extensions to the Standard Model, axion-like particles (ALPs) occupy a central place, offering a compelling solution to some of these outstanding puzzles.  
Motivated by the solution of the strong CP problem~\cite{Peccei:1977hh,Peccei:1977ur,Wilczek:1977pj,Weinberg:1977ma}, ALPs are now present in many extensions of the Standard Model (SM), particularly in scenarios in which a global symmetry is spontaneously broken by a new confining sector, thus yielding Goldstone bosons that behave as ALPs. See for example Ref.~\cite{Choi:2020rgn} for an extended recent review. 

Although traditionally most studies revolved about the coupling of light ALPs to photons and electron-positron pairs, along with their cosmological, astrophysical and detector signatures, there has been quite a lot of recent activity in obtaining constraints on the ALP coupling to other SM particles, by exploring 
novel signatures at colliders. By examining the various channels through which ALPs may be produced, the aim is to identify distinctive signatures that could signal the presence of these elusive particles amidst the background of SM interactions.
For example, the ALP coupling to gluons was explored in Refs.~\cite{Mimasu:2014nea,Ghebretinsaea:2022djg}, the coupling to diboson pairs was considered in~\cite{Jaeckel:2015jla,Brivio:2017ije, Bauer:2017ris,Craig:2018kne} and more recently the ALP coupling to top quarks has been studied in detail in~\cite{Esser:2023fdo,Phan:2023dqw,Blasi:2023hvb,Rygaard:2023dlx}. The collider probes have demonstrated that the mass reach of ALP searches at colliders is quite broad, going beyond pure resonant signatures and involving a variety of non-resonant probes~\cite{Gavela:2019cmq,No:2015bsn,Carra:2021ycg, Biswas:2023ksj}, and this motivated further collider studies. 

One of the couplings that is still relatively unexplored is the coupling of ALPs to Higgs. In Ref.~\cite{Anisha:2023ltp} the coupling of ALPs to the Higgs boson was sifted by means of an Effective Field Theory (EFT) formalism. The Higgs EFT (HEFT) Lagrangian was augmented by a leading order EFT chiral ALP Lagrangian, in which the ALP couples to the SM states~\cite{Brivio:2017ije}. These interactions introduce modifications to the Higgs boson propagation and decay in HEFT, by implying modifications to the Higgs decay rate, as well as to the theoretical predictions for multi-Higgs and multi-top final states. The study shows the importance of combining the traditional Higgs strength measurement with observations of di-Higgs and four-top final states, as these provide significant sensitivity to the ALPs coupling to the Higgs.

\begin{figure}[ht!]
    \centering
    \begin{tikzpicture}[scale=1]
     \draw[decoration={aspect=1, segment length=1.8mm, amplitude=0.7mm,coil},decorate] (-1,1) -- (0,0)-- (-1,-1);
     \draw[dashed] (0,0)-- (1,0);
     \draw[decoration={aspect=0, segment length=1.8mm, amplitude=0.7mm,coil},decorate] (1,0) -- (2,-1);
     \draw[dashed] (2,1)-- (1,0);
     \draw[dashed] (1,0)-- (2,0);
     \draw (2.25,0) node {$h$};
     \draw (2.25,1) node {$h$};
     \draw (1.01,-.01) node {$\bullet$};
     \draw (.01,-.01) node {$\bullet$};
     \draw (2.25,-1) node {$Z$};
     \draw (.5,.25) node {$a$};
     \draw (-1.25,-1) node {$g$};
     \draw (-1.25,1) node {$g$};
\end{tikzpicture}
    \caption{Feynman diagram for di-Higgs production with an associated $Z$ boson via a non-resonant ALP.}
    \label{fig:feynman_diagram_DiHiggsALP}
\end{figure}
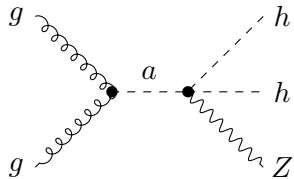

In this work we adopt a different approach and focus on di-Higgs production with an associated $Z$ boson mediated by a non-resonant ALP, as in Fig.~\ref{fig:feynman_diagram_DiHiggsALP}, a process that can be described by a Chiral-ALP EFT~\cite{Brivio:2017ije}. 

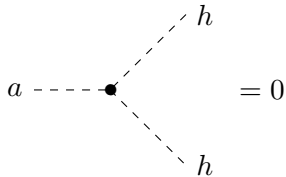
\begin{figure}[ht!]
\centering
\begin{tikzpicture}[scale=1]
     \draw[dashed] (0,0)-- (1,0);
     \draw[dashed] (2,1)-- (1,0);
     \draw[dashed] (1,0)-- (2,-1);
     \draw (2.25,1) node {$h$};
     \draw (2.25,-1) node {$h$};
     \draw (1.01,-.01) node {$\bullet$};
     \draw (-.25,0) node {$a$};
     \draw (3,0) node {$=0$};
\end{tikzpicture}
    \caption{The CP-conserving ALP coupling to two SM Higgs bosons vanishes.}
    \label{fig:ahhzero}
\end{figure}

It is important to note that the inclusion of a $Z$-boson in the final state is essential to achieve a non-zero amplitude. Since the axion-like particle (ALP) is a pseudo-scalar ($0^-$), the combined quantum numbers of the final state must also be pseudo-scalar. With only two identical Higgs bosons in the final state, as depicted in Fig.~\ref{fig:ahhzero}, the wavefunction exhibits even parity, resulting in a state of $0^+$, which is prohibited. However, the presence of a $Z$ boson in the final state can yield a $0^-$ state, allowing only the longitudinal polarization with a zero spin projection  and odd parity to contribute. In App.~\ref{app:FR}, we demonstrate this explicitly by deriving the Feynman rules and calculating the amplitude for on-shell final state Higgs and $Z$ bosons. 

This paper is organised as follows: In Sec.~\ref{sec:chiral_ALP} we briefly review the underlying Chiral-ALP EFT, in Sec.~\ref{sec:dihiggs} we calculate the cross section for the process $gg \rightarrow a \rightarrow h h Z$. In Sec.~\ref{sec:diHiggs-search} we use Run 2 LHC data for di-Higgs events to put constraints on the ALP-Higgs coupling for 3 benchmark scenarios, and assuming the $Z$ decays are not vetoed. We also show the ALP signal for three differential distributions and claim that a dedicated analysis of such distributions could further enhance the sensitivity to the ALP signal. 
In Sec.~\ref{sec:smoking} we propose a tailored search for di-Higgs and detectable $Z$ final states. In Sec.~\ref{sec:tops} we compare the ALP coupling to Higgs and $Z$ in the Chiral-ALP EFT with the loop-induced couplings in a linear ALP EFT that can be mediated via top quark loops. Finally in Sec.~\ref{sec:conclusion} we conclude.

\section{The bosonic chiral ALP}
\label{sec:chiral_ALP}

In this section we review the theory of an ALP coupled to the SM with a generic electroweak symmetry breaking (EWSB), a theory that was developed in Ref.~\cite{Brivio:2017ije}. 

Initially, we examine the linear ALP theory, a scenario in which, at leading order in an expansion in terms of the inverse of the ALP scale $f_a$, we focus on the dimension-five interactions with the SM gauge fields, 
\begin{equation}
    {\cal L} \supset - \frac{a}{f_a} \left(c_{\tilde B} B_{\mu\nu}\tilde B^{\mu\nu}+ c_{\tilde W} W^a_{\mu\nu}\tilde W^{a,\mu\nu}+c_{\tilde G} G^a_{\mu\nu}\tilde G^{a,\mu\nu}\right) \ .
\end{equation}
Additionally, one may include the coupling to the Higgs doublet, $H$,
\begin{equation}\label{ec:linear}
    {\cal O}_{aH} = i \left( H^\dagger \overleftrightarrow{D}_\mu H\right) \, \frac{\partial^\mu a}{f_a} \ ,
\end{equation}
which, after a field redefinition of the Higgs field, leads to fermionic couplings proportional to the Yukawas,
\begin{equation}
     {\cal O}_{a\Psi} = i \frac{a}{f_a} \, (\bar Q_L Y_U \tilde H u_R + \ldots)+ h.c.
\end{equation}
On the other hand, in the HEFT framework, the Higgs particle is a singlet, and it supports a wider family of UV theories than SMEFT, which assumes the doublet structure $H$ (see \textit{e.g.}~\cite{Alonso:2016oah,Cohen:2020xca,Gomez-Ambrosio:2022qsi,Gomez-Ambrosio:2022why,Delgado:2023ynh} for the formal and experimental distinction of these two theories). This means that the Higgs boson can couple to other particles through polynomial \textit{flare} functions such as
\begin{equation}\label{ec:poly}
    {\cal F}_i(h) = 1 + a_i \, \frac{h}{v} + b_i \, \left(\frac{h}{v}\right)^2+ \ldots \ ,
\end{equation}
where $i$ denotes a generic operator. Within HEFT, we find one additional operator at leading order,
\begin{equation}
\mathcal{L}_{\rm chiral}^{LO} = \frac{1}{2} \left(\partial_{\mu} a \right) \left( \partial^{\mu} a \right) + c_{2D} \mathcal{O}_{2D},
\end{equation}
and, expanding the linear couplings in Eq.~\eqref{ec:linear}, 17 additional possibilities for coupling the ALP to SM fields at NLO,
\begin{equation}
    {\cal L}_{\rm chiral}^{NLO} = \sum_{i=\tilde B, \tilde W, \tilde G} \, c_i \, {\cal O}_{i} + \sum_{j=1}^{17} c_j  \, {\cal O}_{j} \, ,
\end{equation}
where each of the operators $\mathcal{O}_{2D}$,  $\mathcal{O}_1$ - $\mathcal{O}_{17}$ is proportional to a Higgs flare function of the form of Eq.~\eqref{ec:poly}.
The complete basis for these interactions is given in Ref.~\cite{Brivio:2017ije}. Here we will only discuss examples of the operators relevant for the phenomenology of di-Higgs production in Sec.~\ref{sec:dihiggs}. A full list of the operators that enter the Feynman rules discussed in the following is provided in Appendix~\ref{app:operators}. \\
Particularly, at leading order one finds a new type of coupling $c_{2D}$, which induces a two-point function of the form $Z^\mu \partial_\mu a$. This term contributes to the longitudinal component of the $Z$ boson, similarly to the {\it would-be} Goldstone bosons in the SM EWSB scenario. Other relevant operators include terms with different structures, such as the operator $c_{17}$ which is $V_{\mu}\partial^{\mu} \frac{\Box a}{f_a}$.\\
In this framework, ALP couplings to two Higgses are possible, as long as a $Z$ or a $\gamma$ are attached to the vertex, as we will show explicitly in Sec.~\ref{sec:dihiggs}. Note that for this vertex to exist at tree-level, a non-linear realization of EWSB is required.
However, as we will explore in Sec.~\ref{sec:tops}, this vertex can also arise from a linear EWSB scenario via a loop involving top quarks.   

The HEFT expansion (organized w.r.t. the chiral dimension of the operators or number of derivatives in the chiral limit~\cite{Buchalla:2013eza}) is usually believed to be reliable up to $\sqrt{s}=4\pi v$. Nonetheless, as several works have proven (see for example Refs.~\cite{Alonso:2015fsp,Alonso:2016oah}), the combination of the HEFT contributing diagrams, for example for $\omega \omega \to hh$ at LO, where the $\omega$s are the Goldstone Bosons (GBs) of the EWSB and the flare function 
\begin{equation*}
\mathcal{F}_{\text{GB}}(h)=1 + a_{\text{GB}} \, \frac{h}{v} + b_{\text{GB}} \, \left(\frac{h}{v}\right)^2+ \ldots \;,
\end{equation*}
coupling an arbitrary number of Higgs fields to the kinetic term of the GBs, produces the amplitude \cite{Delgado:2014dxa} \begin{equation}\mathcal{A}_{\omega\omega\to hh}\frac{(b_{\text{GB}}-a_{\text{GB}}^2) s}{16\pi^2 v^2}.
\end{equation}
Hence, the cutoff is actually given by
\begin{equation}
    \Lambda^2 = \frac{16\pi^2 v^2}{b_{\text{GB}}-a_{\text{GB}}^2},
\end{equation}
with $a_{\text{GB}}=1$ and $b_{\text{GB}}=1$ in the SM. More formally, the works in the literature show that the cutoff is given by 
\begin{equation}
\Lambda^2 = \frac{16\pi^2 v^2}{R},
\end{equation} 
where $R$ is the scalar curvature of the scalar sector (Higgs and GBs). This curvature is zero in the SM, rendering it valid at all scales, whereas in more generic cases, this curvature will depend on the specific values of the parameters considered. For the case of the ALP, the discussion follows the same line, as this ALP would be  enlarging the scalar sector (it is just an extra pseudo GB, although for a different symmetry). As we will see in Eq.~(\ref{eq:shat}) below, the cutoff of the HEFT+ALP EFT
will be proportional to $f_a v$ divided by the relevant combination of Wilson coefficients.

%%%%%%%%%%%%%%%%%%%%%%%%%%%%%%%%%%%%%%%%%%%%%%%%%%%%%%%%%%%%%%%%%%%%%%%%%%%%%%%%%%%
\section{Di-Higgs in the chiral ALP}\label{sec:dihiggs}
After setting the theoretical framework for the ALP interactions with the SM particles, we move on to discussing the di-Higgs signature. As discussed in Sec.~\ref{sec:intro}, the ALP can couple  to two Higgses in association with a $Z$ or $\gamma$ boson. Currently, there are no searches for a di-Higgs and photon or $Z$ final state, so in Sec.~\ref{sec:diHiggs-search} we will re-cast a current di-Higgs search to place limits on the ALP production. In Sec.~\ref{sec:smoking}, we will propose a dedicated search where the two Higgses are produced in association with a boosted $Z$.

First, as the ALP couples to fermions with a coupling proportional to their mass, the main production mechanism at the LHC is either vector-boson fusion or gluon fusion. The gluon fusion production is shown in Fig.~\ref{fig:feynman_diagram_DiHiggsALP}, where the ALP is produced off-shell and leads to a di-Higgs plus $Z$ final state.
In Ref.~\cite{Gavela:2019cmq}, it was shown that the off-shell production mechanism is not suppressed due to the derivative nature of the ALP-gluon coupling. Instead, the partonic cross section scales as
\begin{equation} \label{eq:shat}
    \hat \sigma (g g \to a \to h \, h \, Z) \propto \hat s^3 \, \frac{c_{\tilde G}^2 \, c_{3B}^2}{v^4 f_a^4} \, ,
\end{equation}
where we have defined $c_{3B}$ as the generic coupling of the ALP to three bosons. 

The specific form of the cross section will depend on the Lorentz structure of $c_{3B}$, which exhibits a rich structure in terms of energy-momentum dependence of the ALP and  bosons. This can be seen in the Feynman rules listed in Fig.~\ref{fig:FR_ahhZ}. 

\begin{figure*}[t!]

\hspace{-1.5cm}
  \begin{subfigure}[b]{0.4\textwidth}
            
    \centering
    \begin{tikzpicture}[scale=1]
     \draw[decoration={aspect=1, segment length=1.8mm, amplitude=0.7mm,coil},decorate] (-1,1) -- (0,0)-- (-1,-1);
     \draw[dashed] (0,0)-- (1,0);
     \draw (.01,-.01) node {$\bullet$};
     \draw (1.25,0) node {$a$};
     \draw (-1.25,-1) node {$g^\mu$};
     \draw (-1.25,1) node {$g^\nu$};

\end{tikzpicture}
\end{subfigure}
\begin{subfigure}[b]{0.475\textwidth}
            
    \centering
\begin{equation*}
\vspace{.8cm}
     -\frac{4i}{f_a}c_{\tilde{G}} \,p_{g1\,\alpha}p_{g2\,\beta}\,\varepsilon^{\mu\nu \alpha \beta}
\end{equation*}   
\end{subfigure}

\hspace{-1.5cm}
\begin{subfigure}[b]{0.4\textwidth}
            
    \centering

        \begin{tikzpicture}[scale=1]
     \draw[dashed] (0,1)-- (1,0);
     \draw[decoration={aspect=0, segment length=2.5mm, amplitude=0.5mm,coil},decorate] (1,0) -- (0,-1);
     \draw[dashed] (2,1)-- (1,0);
     \draw[dashed] (1,0)-- (2,-1);
     \draw (2.25,1) node {$h$};
     \draw (2.25,-1) node {$h$};
     \draw (1.01,-.01) node {$\bullet$};
     \draw (-.25,-1) node {$Z^\mu$};
     \draw (-.25,1) node {$a$};
\end{tikzpicture}
\end{subfigure}
 \begin{subfigure}[b]{0.475\textwidth}
            
    \centering
\begin{align*}
\vspace{.8cm}
    &\frac{g}{8\pi^2 c_{W}v^2f_a}\Big[p_{hh}^\mu (p_a^2\tilde{b}_{11}+p_a\cdot p_{hh}\tilde{b}_{14})+p_a^\mu(p_{hh}^2\tilde{b}_{13}+p_a\cdotp p_{hh}\tilde{b}_{12})\\
    &+2\tilde{a}_{16}(p_{h1}^\mu p_a\cdot p_{h2}+p_{h2}^\mu p_a\cdot p_{h1})+4\tilde{a}_{15} p_a^\mu p_{h1}\cdot p_{h2}\\
    & -p_a^\mu(16\pi^2 v^2 \tilde{b}_{2D}-\tilde{b}_{17}p_a^2)+ 2\pi s_{2 W} \tilde b_{310} (p_{Z}^2 p_a^\mu -p_{Z}^\mu p_a\cdot p_Z)/e\Big]
\end{align*}
\end{subfigure}

\hspace{-1.5cm}
  \begin{subfigure}[b]{0.4\textwidth}
            
    \centering

        \begin{tikzpicture}[scale=1]
     \draw[dashed] (0,1)-- (1,0);
     \draw[decoration={aspect=0, segment length=2.5mm, amplitude=0.5mm,coil},decorate] (1,0) -- (0,-1);
     \draw[dashed] (2,1)-- (1,0);
     \draw[dashed] (1,0)-- (2,-1);
     \draw (2.25,1) node {$h$};
     \draw (2.25,-1) node {$h$};
     \draw (1.01,-.01) node {$\bullet$};
     \draw (-.25,-1) node {$A^\mu$};
     \draw (-.25,1) node {$a$};
\end{tikzpicture}
\end{subfigure}
\begin{subfigure}[b]{0.475\textwidth}
            
    \centering
\begin{equation*}
\vspace{.8cm}
\frac{1}{2\pi v^2 f_a} \left(\tilde{b}_3 c_W+\tilde{b}_{10} s_W \right)(p^\mu_\gamma p_a^2-p^2_\gamma p_a^\mu)\stackrel{\tiny\text{on shell}}{=}0
\end{equation*}   
\end{subfigure}
\caption{Feynman rules for the ALP coupling to gluons (first row), and on-shell Higgses with a $Z$ (second row) or a photon (third row). Cf.\ App.\ \ref{app:operators} for a definition of the relevant operator structures.}
    \label{fig:FR_ahhZ}
\end{figure*}

In the first row of Fig.~\ref{fig:FR_ahhZ}, we show the momentum dependence of the gluon-ALP coupling, with the characteristic $\varepsilon^{\mu\nu\alpha\beta}$ structure stemming from the $a \, G^a_{\mu\nu} \, \tilde G^{\mu\nu,a}$ interaction term. 
The second row shows  all structures for the ALP-diHiggs-$Z$ vertex spanned by the chiral ALP Lagrangian described in Ref.~\cite{Brivio:2017ije}, where we have already assumed the Higgses and $Z$ boson in the final state are on-shell, i.e. $p_{h,Z}^2=m_{h,Z}^2$, and have defined the di-Higgs combination of momenta as $p_{hh}=p_{h_1}+p_{h_2}$.  Here $s_W$, $c_W$ are defined as the sine and cosine of the Weinberg angle. We have also defined $\tilde b_{310}=\tilde{b}_{3}s_{W}-\tilde{b}_{10} c_{W}$ with respect to the original notation in Ref.~\cite{Brivio:2017ije}. Note that the coupling $\tilde c_{2D}$ has a slightly different momentum structure compared to the others, which would change the scaling in Eq.~\eqref{eq:shat} to $\hat s/f_a^4$ instead of $\hat s^3/f_a^4 v^2$.

Finally, in the third row we show the ALP-diHiggs-photon coupling, which vanishes when the $\gamma$ is on-shell. Indeed, for an on-shell photon the second term vanishes as $p_\gamma^2=0$. Moreover, the first term also vanishes once contracted with the polarization vector, $\varepsilon^T_\mu p_\gamma^\mu=0$, since the photon polarizations are transverse. 

In Appendix~\ref{app:FR} we present the full Feynman rule for the ALP-diHiggs-$Z$, as well as a discussion of the kinematics associated with this momentum dependence.

To summarize this section, we consider the off-shell production of an ALP through gluon fusion, leading to two Higgses and a $Z$ boson, as shown in Fig.~\ref{fig:feynman_diagram_DiHiggsALP}. This process depends on the coupling $c_{\tilde G}$ and a combination of the many possible $ahhZ$ couplings shown in Fig.~\ref{fig:FR_ahhZ}. To express our limits in terms of the scale $f_a$ and the coupling coefficients $c_{3B}$, we will choose three different benchmarks, 
\begin{eqnarray}
   & &  \textrm{ Benchmark 1: } \tilde{b}_{3,10-17}= \tilde{a}_{15,16} = \tilde b_{2D} = 1 \ ,\nonumber \\ & &  \textrm{ Benchmark 2: }  \tilde b_{2D} = 1 \textrm{, all others set to zero, }\nonumber \\
     & &  \textrm{ Benchmark 3: } \tilde{b}_{17} = 1 \textrm{, all others set to zero. } \label{eq:bench}
\end{eqnarray}

Benchmark 1 represents a situation where all the couplings are switched on with similar strengths. Benchmarks 2 and 3 isolate the effect of two types of couplings with specific behaviour. The coupling $\tilde b_{2D}$ represents a $Z-$axion mixing and is un-suppressed in the chiral expansion discussed  in Ref.~\cite{Brivio:2017ije}.  On the other hand, the coupling $\tilde b_{17}$ is directly proportional to $p_a^2$, the $\hat s$ of the event and a direct test of the validity of the EFT expansion. See App.~\ref{app:FR} for more details on the kinematic behaviour of these couplings.

%%%%%%%%%%%%%%%%%%%%%%%%%%%%%%%%%%%%%%%%%%%%%%%%%%%%%%%%%%%%%%%%%%%%%%%%%%%%%%%%%%%
\section{ALP contribution to di-Higgs in the two $b$ and two photon final state}
\label{sec:diHiggs-search}

In this section we use current Run 2 LHC data to place limits on the ALP-mediated process of Fig.~\ref{fig:feynman_diagram_DiHiggsALP}.

There is currently no search for two Higgses in association with a vector boson, but one can use the di-Higgs searches as long as the final state $Z$ products are not vetoed in the analysis.  
To illustrate this procedure, we consider the following search, 
\begin{equation}
    p p \to h h + X \to b \, \bar b \, \gamma \, \gamma \, + X \ ,
\end{equation}
where the Higgses decay into a pair of $b$-jets and pair of photons. There are searches both from CMS~\cite{CMS:2022gjd} and ATLAS~\cite{ATLAS:2023gzn}, and we will focus on the ATLAS analysis as they provide interesting distributions with strong sensitivity to the ALP production. The CMS analysis provides the \hepdata\footnote{See \url{https://www.hepdata.net/record/ins2081829}.} but the information is encoded in terms of BDT variables which we could not directly recast for our case.~\footnote{There would be other relevant searches: CMS $hh\to b\bar{b}W^+W^-$~\cite{CMS:2023qiw}, CMS  $hh\to b\bar{b}\tau^+\tau^-$~\cite{CMS:2022hgz}, CMS $hh\to b\bar{b}\gamma\gamma$~\cite{CMS:2020tkr}, ATLAS $hh\to b\bar{b}b\bar{b}$~\cite{ATLAS:2020jgy}, CMS $hh\to b\bar{b}b\bar{b}$ \cite{CMS:2022cpr}, ATLAS  $hh\to b\bar{b}b\bar{b},\, b\bar{b}\tau^+\tau^-,\, b\bar{b}\gamma\gamma$~\cite{ATLAS:2022kbf}, ATLAS  $hh\to b\bar{b}b\bar{b}$~\cite{ATLAS:2023qzf}, CMS  $hh\to b\bar{b}b\bar{b}$~\cite{CMS:2022gjd}, and CMS (all channels)~\cite{CMS:2022dwd}. However their inclusion goes beyond the scope of this paper.}

Focusing on the ATLAS analysis~\cite{ATLAS:2023gzn}, we will discuss the limit one can extract from the  signal region ($m_{\gamma\gamma}\in $ [120,130] GeV)  as well as three interesting distributions displayed in the auxiliary material:~\footnote{ See ~\url{https://atlas.web.cern.ch/Atlas/GROUPS/PHYSICS/PAPERS/HDBS-2021-10/}.} the di-Higgs invariant mass distribution $m_{hh}$, the transverse sphericity $S_T$ and the $\Delta R$ between the two photons into which one Higgs decays, $\Delta R_{\gamma\gamma}$.

We  consider the ALP-mediated production of two Higgses in this final state $b\bar b \gamma \gamma$ plus a $Z$ boson decaying to neutrinos or jets. These events will typically enter into the di-Higgs signal, as no explicit vetoes have been placed on missing energy, and the veto on additional jets is quite lax. Indeed, this analysis vetoes on six or more central jets, defined as jets with $p_T^j>$ 25 GeV inside the inner detector, $|\eta|_j$< 2.5. Therefore, ALP-mediated events with $Z$ decaying into neutrinos or into less than 6 central jets would typically pass the selection criteria for di-Higgs. 

We  simulate the ALP-mediated contribution using the Monte-Carlo event generator \madgraph~\cite{Alwall:2014hca} and the {\tt UFO}~\cite{Darme:2023jdn} model based on Ref.~\cite{Brivio:2017ije}.~\footnote{The UFO can be downloaded from the Feynrules webpage, the {\it ALP-chiral} model in \url{http://feynrules.irmp.ucl.ac.be/wiki/ALPsEFT}.} With this simulation, we  compute the  cross section given by
\begin{equation}
  \sigma (p p \to  a^* \to h \, h \, Z) \cdot 2 \cdot BR(h \to b \bar b ) \cdot BR(h \to \gamma \gamma )  \cdot BR(Z \to \textrm{ jets or }\nu\bar \nu) \cdot {\cal A},
\end{equation}
where ${\cal A}$ stands for the acceptance within the selection cuts
\begin{eqnarray}\label{basicuts}
    & & p_T^{\gamma_{1,2}}> 35 \, (25) \textrm{ GeV, and } |\eta_\gamma| < 2.4 \nonumber \\
    & & p_T^b > 25 \textrm{ GeV, and } |\eta_b| < 2.5 \ ,
\end{eqnarray}
as well as a b-tag flat rate of $\epsilon_{b-tag}=$0.77 per $b$-jet. Additionally, we impose isolation criteria for the $b$-jets and photons along the lines of the ATLAS analysis~\cite{ATLAS:2023gzn} with the requirement that $\Delta R$ in between all objects larger than 0.4. Finally, we focus on the {\it high-mass} region, defined as $m_{hh}>$ 350 GeV. 

After imposing all the selection cuts, we obtain an estimate for the ALP-mediated cross section, which parametrically depends on the ALP coupling to the gluon $c_{\tilde G}$, the couplings to the $hhZ$ bosons $c_{3B}$ and the ALP scale $f_a$ as discussed in Eq.~\eqref{eq:shat}. 
We will denote the combination of gluon and electroweak bosons with the letter $c$, namely
\begin{equation}
    c \equiv c_{\tilde G} \, c_{3B} \, ,
\end{equation}
and will show results for $c_{3B}$  chosen along the three benchmarks defined in Eq.~\eqref{eq:bench}.

%%%%%%%%%%%%%% Section 4.1 %%%%%%%%%%%%%%%%%%%%%%%%%%%
\subsection{Limits from the \textit{high-mass} region}
\label{subsec:totxsec}
In this section we place limits on the three benchmarks using the number of observed events and the SM expectation provided by ATLAS in the high-mass signal region, from Table 3 of~\cite{ATLAS:2023gzn}.

To test the normalization of our Monte-Carlo simulation, we compute the $\gamma\gamma b \bar b$ SM background with \madgraph~\cite{Alwall:2014hca},
and compare our background estimation with the total number of events in that table. Applying the basic cuts in Eq.~\eqref{basicuts} and the b-tagging flat rate, and focusing in the region $m_{\gamma\gamma}\in [120, 130]$ GeV and $m_{b\bar b}\in [105,160]$ GeV, we obtain
$\sigma_{\rm continuum} \simeq 0.52$ fb, which leads to approximately 73 events, to be compared with the 76 events reported in Table 3 of Ref.~\cite{ATLAS:2023gzn} combining the high- and low-mass regions. Therefore, the Monte Carlo simulation of the continuous background provides a good approximation of the SM events.
 
We now  compare the total number of measured events ($n_{\rm obs}$) in the entire \textit{high-mass} region to the background estimate provided by ATLAS $n_{\rm BG}$, and our signal prediction $n_{\rm s}$, which depends on the combination $c/f_a^2$. 
\begin{figure}[ht!]
    \centering
\includegraphics[width=.48\columnwidth]{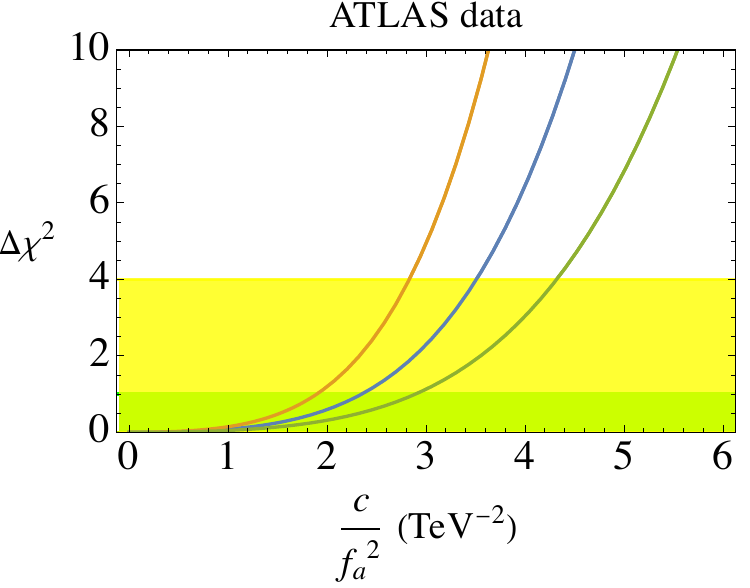}
    \caption{Dependence of $\Delta \chi^2$ on $c/f_a^2$.  The three lines correspond to the benchmarks described in Eq.~\eqref{eq:bench}, Benchmark 1 (blue line), Benchmark 2 (orange line) and Benchmark 3 (green line).}
    \label{fig:chi2_HM}
\end{figure}

We perform a $\chi^2$ test 
\begin{equation}
    \chi^2 \left(\frac{c}{f_a^2}\right)= \left(\frac{n_{\rm obs}-n_{\rm BG}-n_{\rm s}(c/f_a^2)}{\Delta_{\rm BG}}\right)^2 \ ,
\end{equation}
where $\Delta_{\rm BG}$ denotes the quadratic sum of the background uncertainties in the three \textit{high-mass} regions as provided by ATLAS. 
Note that no uncertainties on the measurement are provided and so here we assume the background uncertainty to be the dominant contribution to the total uncertainty.

Fig.~\ref{fig:chi2_HM} shows $\Delta \chi^2 = \chi^2(c/f_a^2) - \chi^2_{\rm min}$ for the three benchmarks. The limits corresponding to 2 standard deviations translate into
\begin{equation} 
\label{eq:HMlim}
    f_a > (0.53, 0.59, 0.48) \times \sqrt{c} \textrm{ TeV} 
\end{equation}
for Benchmark 1, Benchmark 2 and Benchmark 3, respectively. 

For completeness, if instead we took into account only the experimental error provided by the statistical uncertainty, given by $\Delta_{\rm exp} = \sqrt{N_{\rm events}} \approx 4.12$, we find the following weaker limits on $f_a$:
\begin{equation} 
    f_a > (0.40, 0.44, 0.36) \times \sqrt{c} \textrm{ TeV.} 
\end{equation}
We now move on to describe the differential distributions which could be used by the experiments to further improve our estimate in Eq.~\ref{eq:HMlim}.

%%%%%%%%%%%%%% Section 4.2 %%%%%%%%%%%%%%%%%%%%%%%%%%%
\subsection{ALP-mediated di-Higgs differential distributions}
\label{subsec:mhh}
Although the ATLAS analysis provides additional material with interesting differential distributions, these are normalized to 1 and focused on regions which are defined by BDT classifiers which cannot be reproduced with our simulation tools,  leaving insufficient details for comparing our signal events with the observed data.  Therefore, in this section we will simply describe the behaviour of the ALP-mediated production  to motivate further analyses using these distributions. 

We start with the di-Higgs invariant mass. The parton-level growth shown in Eq.~\eqref{eq:shat} translates into a distribution  similar to what one would expect from an EFT modifier $\kappa_{2V}$. However, in contrast to the EFT case analyzed by ATLAS, the ALP-mediated signal does not exhibit interference with the SM di-Higgs production, as discussed in the Sec.~\ref{sec:intro} and depicted in Fig.~\ref{fig:ahhzero}.  

\begin{figure}[t!]
    \centering
         \centering         \includegraphics[width=.47\textwidth]{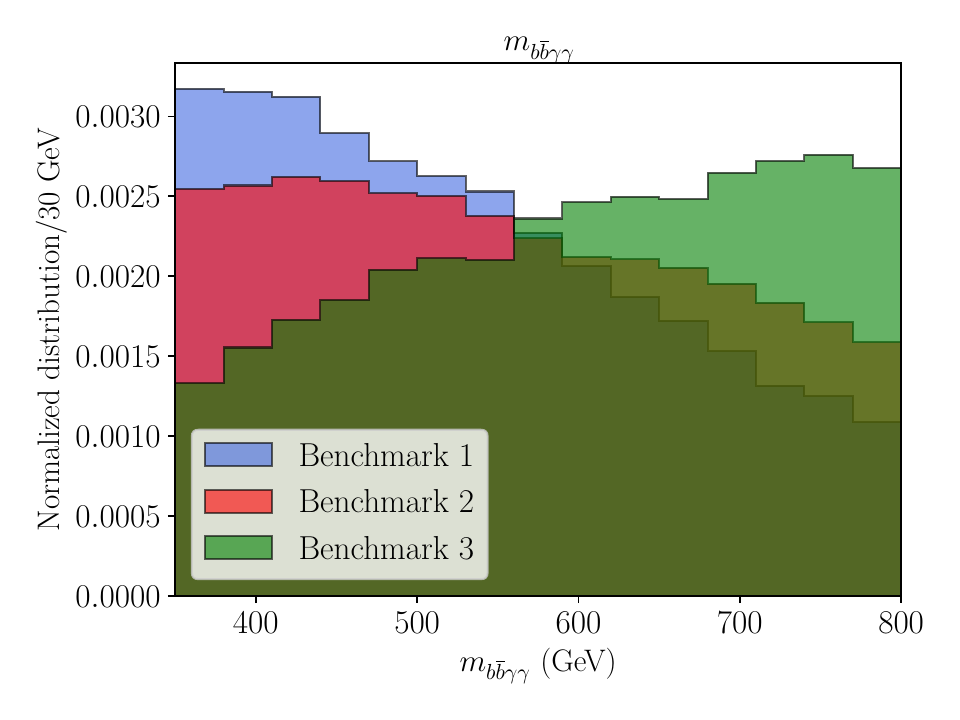}\\
         
         \includegraphics[width=.47\columnwidth]{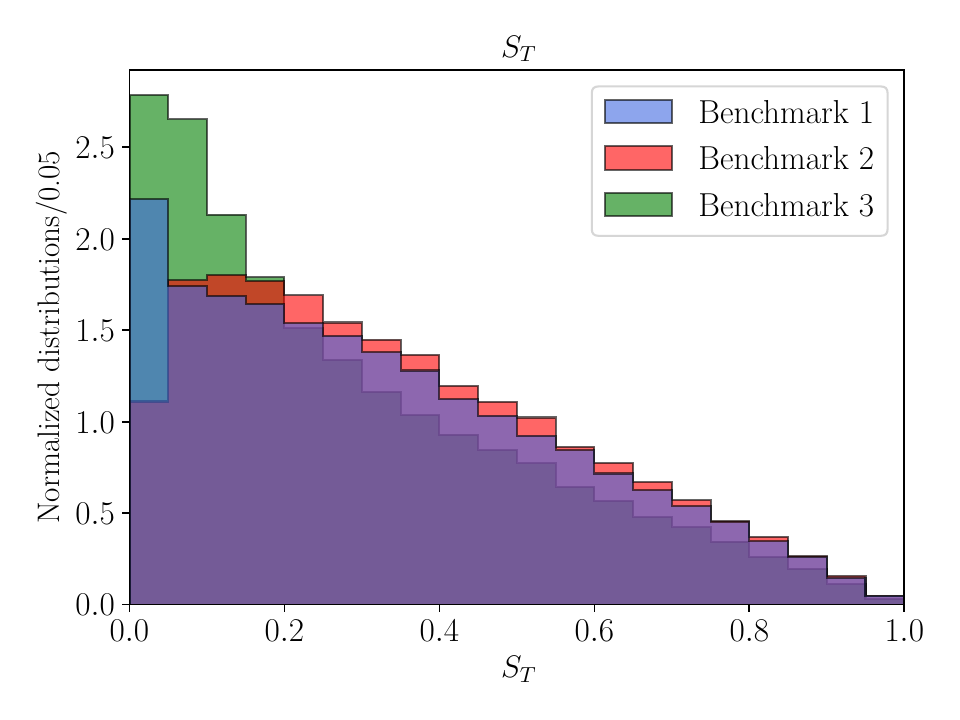}
         \includegraphics[width=.47\columnwidth]{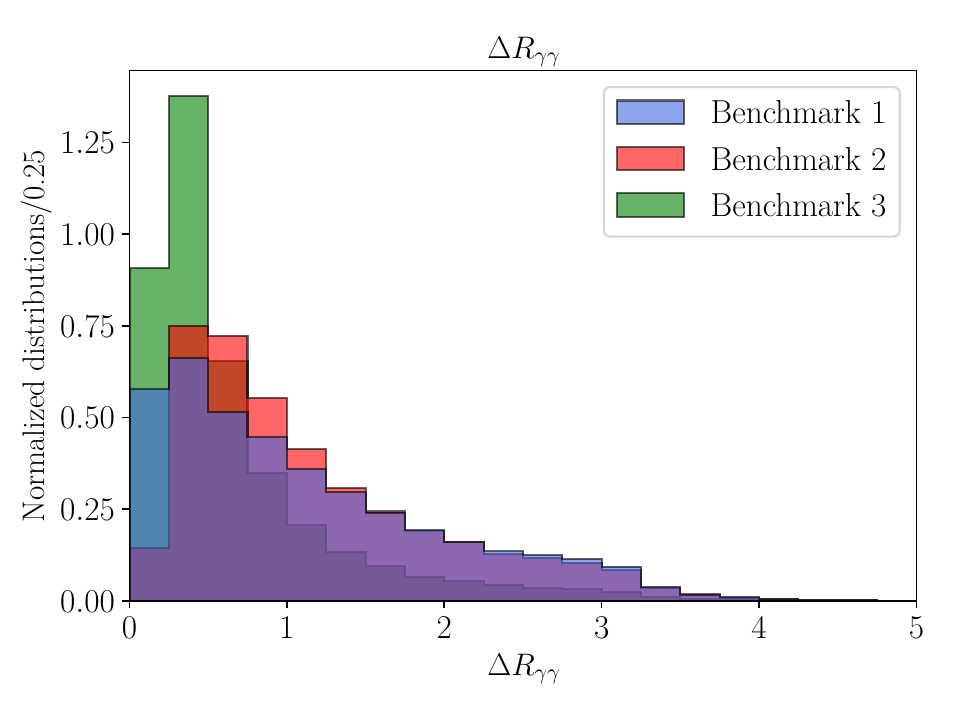}
         
          \caption{{\it Upper:} normalised distribution of the invariant mass distributions of the two Higgs products, shown for the three benchmarks in Eq.~\ref{eq:bench}.  {\it Lower left:} normalised transverse sphericity $S_T$ distribution and {\it lower right:} normalised $\Delta R_{\gamma \gamma}$ distribution.}
        \label{fig:mhh}
\end{figure}

In the upper panel of Fig.~\ref{fig:mhh} we show the normalized distribution of the di-Higgs system invariant mass, $m_{hh}$, for the three benchmark choices. The distribution in the case of Benchmark 1 exhibits the mildest behaviour of the three benchmarks, although we have checked that it does grow faster than the SM continuum. As expected, Benchmark 3, based on the operator $\tilde b_{17}$, exhibits the strongest growth. See App.~\ref{app:FR} for more details.

We now explore the transverse sphericity, $S_T$, which quantifies how symmetric the scattering event is with respect to the beam axis. The transverse sphericity is defined in terms of the eigenvalues, $\lambda_1\geq\lambda_2$, of the linearised version of the transverse momentum tensor~\cite{Banfi:2010xy}
\begin{align}
\boldsymbol{S_{XY}^L}=\frac{1}{\sum_i p_T^{(i)}}\sum_i \frac{1}{p_T^{(i)}} \begin{pmatrix}
    {p_x^{(i)\,2}} & p_x^{(i)}p_y^{(i)}\\
    p_y^{(i)}p_x^{(i)} & p_y^{(i)\,2}
\end{pmatrix}
\end{align}
where $p_T^{(i)}=|(p_x^{(i)},p_y^{(i)})|$ is the transverse momentum of the $i$-th particle in the final state. The transverse sphericity is given by
\begin{align}
    S_T\equiv \frac{2\lambda_2}{\lambda_1+\lambda_2}\,.
\end{align}
In the lower-left panel in Fig.~\ref{fig:mhh} we plot the normalized $S_T$ distribution from our simulation for Benchmarks 1, 2 and 3.  It is striking that the ALP distribution is concentrated at low values of sphericity, as the products of the ALP-mediated process are very boosted and tend to collimate. Moreover, the $S_T$ calculation includes the two Higgs decay products, and misses the $Z$ part of the event.    

Another distribution provided by ATLAS that can potentially provide interesting information is the $\Delta R_{\gamma\gamma}$ of the decay products of one of the Higgses in the event. The distribution is shown in the right-lower panel of Fig.~\ref{fig:mhh}. Here we see the ALP-mediated process produce collimated photons, as the ALP-mediated process is enhanced at high-momentum transfer, see Appendix~\ref{app:FR}. 
 \noindent

 %%%%%%%%%%%%%%%%%%%%%%%% section 5 %%%%%%%%%%%%%%%%%%%%%%%%%%%%%%
\section{A smoking gun of the ALP-mediated di-Higgs production}
\label{sec:smoking}

In the previous sections, our analysis focused on the ongoing search for di-Higgs events, assuming that the additional $Z$ boson remains untagged, thereby circumventing detection vetoes. Specifically, we assumed that the $Z$ boson decays into hadrons or into neutrinos. 

\begin{figure}[ht!]
    \centering
    \begin{tikzpicture}[scale=1]
     \draw[decoration={aspect=1, segment length=1.8mm, amplitude=0.7mm,coil},decorate] (-1,1) -- (0,0)-- (-1,-1);
     \draw[dashed] (0,0)-- (1,0);
     \draw[decoration={aspect=0, segment length=1.8mm, amplitude=0.7mm,coil},decorate] (1,0) -- (2,-1);
     \draw[dashed] (2,1)-- (1,0);
     \draw[dashed] (1,0)-- (2,0);
     \draw[->-,thick] (3.2,.5)-- (2,0);
     \draw[->-,thick] (2,0) -- (3.2,-.5);
     \draw[->-,thick,rotate=26] (3.2,-.5) -- (2.2,0);
     \draw[->-,thick,rotate=26] (2.2,0) -- (3.2,.5);
     \draw[->-,thick,rotate=-26] (3.2,-.5) -- (2.2,0);
     \draw[->-,thick,rotate=-26] (2.2,0) -- (3.2,.5);
     \draw (1.6,0.25) node {$h$};
     \draw (3.5,0.45) node {$\bar{b}$};
     \draw[rotate=28] (3.5,0.45) node {$\bar{b}$};
     \draw (3.5,-0.45) node {${b}$};
     \draw[rotate=26] (3.5,-0.45) node {${b}$};
     \draw[rotate=-25] (3.5,0.45) node {$l^-$};
     \draw[rotate=-27] (3.5,-0.45) node {$l^+$};
     \draw (1.25,1) node {$h$};
     \draw (1.25,-1) node {$Z$};
     \draw (1.01,-.01) node {$\bullet$};
     \draw (.01,-.01) node {$\bullet$};
     \draw (.5,.25) node {$a$};
     \draw (-1.25,-1) node {$g$};
     \draw (-1.25,1) node {$g$};
\end{tikzpicture}
    \caption{Feynman diagram of the ALP-mediated process $hhZ\to b \bar b b \bar b \ell^+ \ell^-$. }
    \label{fig:HHZto4b2L}
\end{figure}
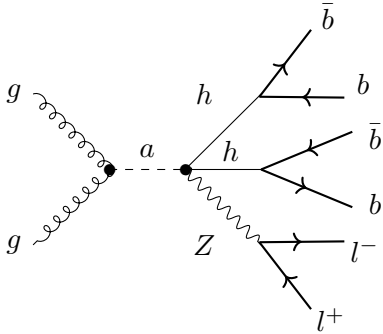

In this section, we propose a new di-Higgs search, coupling it with a leptonic $Z$  boson. The clean leptonic decays facilitate a refined search, wherein the Higgs bosons predominantly decay into their primary channel, $b$-jets. The Feynman diagram depicting the ALP-mediated process under consideration is illustrated in Fig.~\ref{fig:HHZto4b2L}.

\begin{figure}[t!]
    \centering
    \includegraphics[width=.47\columnwidth]{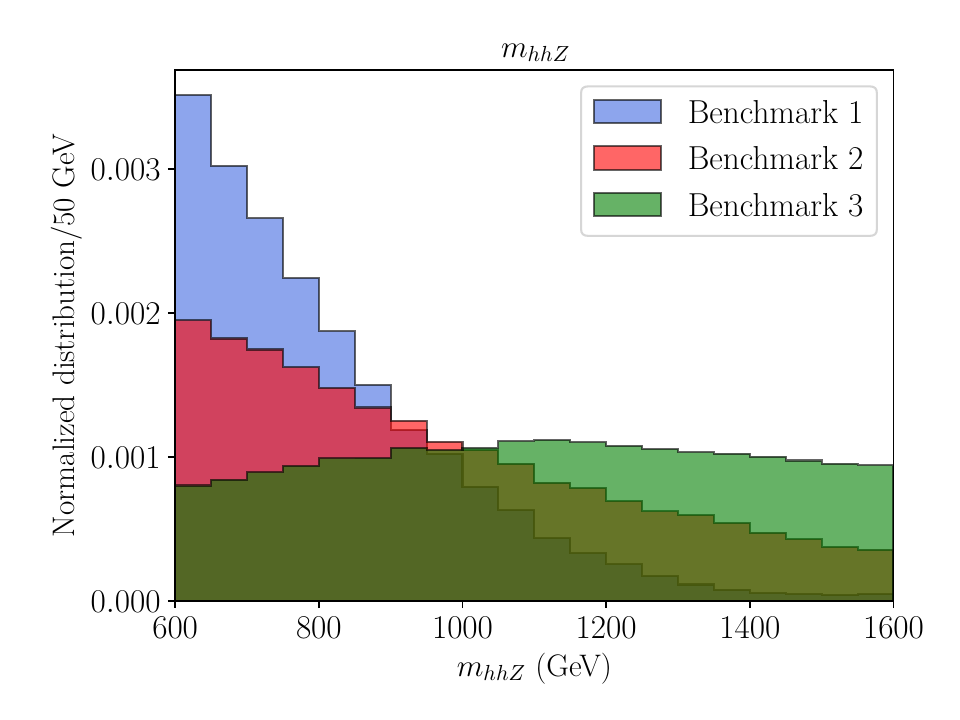}
    \includegraphics[width=.47\columnwidth]{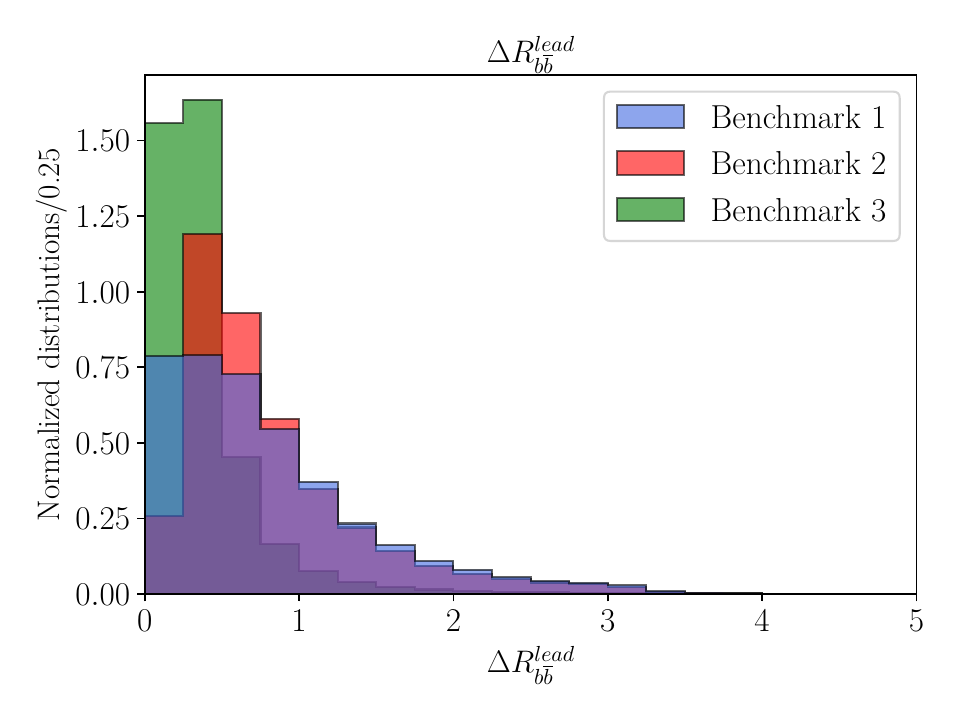}    \includegraphics[width=.47\columnwidth]{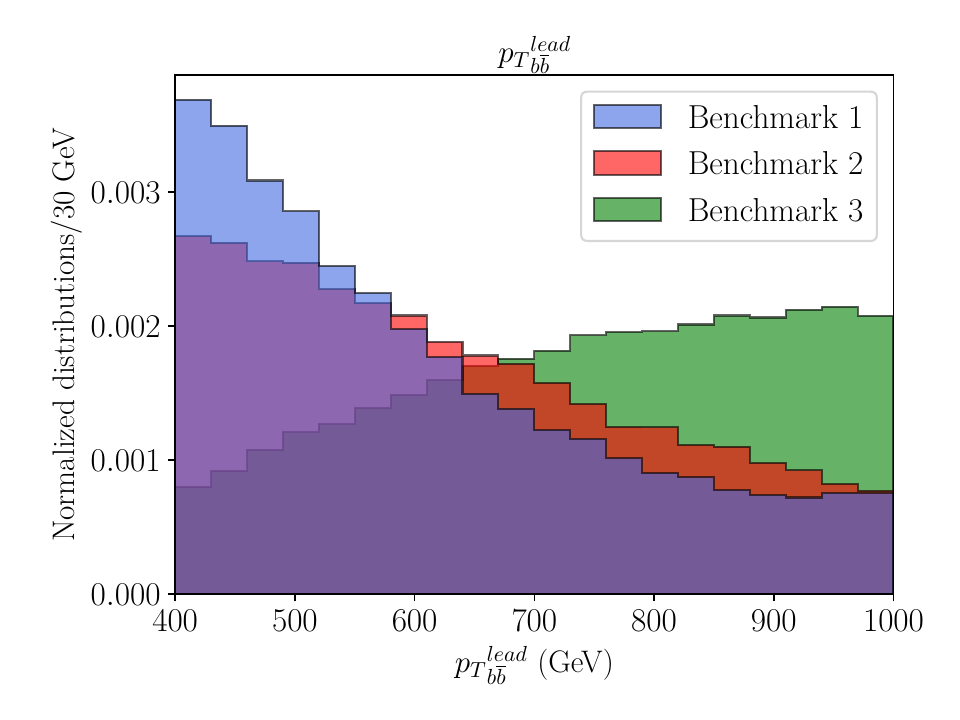}
    \includegraphics[width=.49\columnwidth]{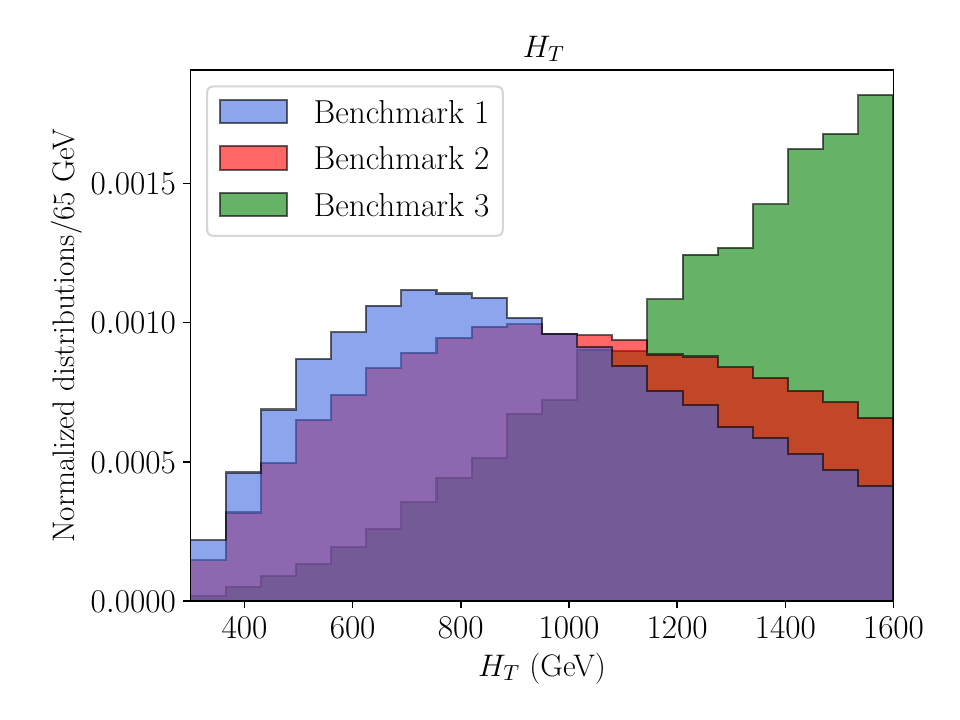}
    \caption{Distributions of the ALP-mediated $hhZ\to 4\, b \ell^+\ell^-$ final state: invariant mass of the system $m_{hhZ}$ (top-left), separation between the b quarks from the leading Higgs $\Delta R_{bb}$ (top-right), transverse momentum of the leading Higgs $p_T^{h}$ (bottom-left) and scalar sum of the $p_T$ of all b quarks (bottom-right). The blue, magenta and green distributions correspond to the three benchmark choices given in Eq.\eqref{eq:bench}. }
    \label{fig:mHHZDRtoleptons}
\end{figure}

Now the whole system of four $b$-jets and two energetic charged leptons can be tagged. These final states are very boosted and collimated, as can be seen from the distributions in Fig.~\ref{fig:mHHZDRtoleptons}. 

In the top left panel of this figure we depict the energy of the event, $m_{hhZ}$, for the three benchmarks discussed in Sec.~\ref{sec:dihiggs}. The case with nonzero $\tilde b_{17}$ (green distribution) exhibits the strongest energy growth, a behaviour we discuss in more detail in App.~\ref{app:FR}. Next comes the magenta distribution, due to a nonzero $\tilde b_{2D}$. Finally, the mildest behaviour comes from the benchmark with all $c_{3B}$ coefficients set to one, shown as a blue distribution, where a partial cancellation between  $\tilde b_{2D}$ and $\tilde b_{17}$ occurs, see App.~\ref{app:FR} for more details. 

Those boosted Higgses and $Z$ boson would  lead to collimated decay products, a fact that can be seen in the top right panel of Fig.~\ref{fig:mHHZDRtoleptons}, where we plot the separation between the two $b$-jets from the leading Higgs, $\Delta R_{bb}$. This distribution is peaked around small values, which will lead to non-isolation of the $b$-jets in many cases, where these energetic and merged $b$-jets will act as fat jets. 

The lower panels in Fig.~\ref{fig:mHHZDRtoleptons} depict the $p_T$ of the leading Higgs and the variable $H_T=\sum_{i=1}^4 p_T^{b_i}$. These distributions show that strong selection cuts on the signal can be applied, e.g. $H_T>500$ GeV, with good acceptance. 

To estimate the background from SM production,   we first compute the production cross section of a $Z$ boson plus four $b$-jets. Imposing the selection cuts on all the jets in transverse momentum $p_T^b> m_h/2$ and isolation $\Delta R_{bb}> 0.4$, we obtain a SM value of order 4 fb. 

Removing the isolation condition, the cross section would increase to 7 fb. However, the merging of $b$-jets would result in a different final state topology, aligning more closely with the one examined in Ref.~\cite{Gouzevitch:2013qca}. In this paper, the authors proposed a tagging algorithm for two Higgses into $b$-jets which could go from the fully resolved $4\,b$ final state, to intermediate fat jet + 2 $b$-jets situation, and reaching the two fat jet case. The overall efficiency of this scaling procedure was estimated to be around 50\% across all the channels.  Finally, applying an additional cut on $H_T> $ 500 GeV brings the cross section from 4 to around 2 fb.

The process $Z$+ 4 jets with cuts  $p_T^j> m_h/2$ and isolation $\Delta R_{jj}> 0.4$ leads to 20 pb.
A cut on $H_T> $ 500 GeV brings this cross section down to 12 pb.
Following the ATLAS analysis~\cite{ATLAS:2023gzn}, we assume mis-tagging rates of quarks into $b$-jets is 1/170 for light flavour quarks and 1/5 for charm quarks~\cite{ATLAS:2022qxm}. Therefore, mis-tagging four light quarks for b-quarks would be suppressed by a factor $10^{-9}$, rendering this background irrelevant. Final states with 4 c-jets, which constitute a small portion of the $Z$+jets background, would be suppressed by a factor of order $2\times 10^{-3}$, again rendering this background negligible respect to the genuine $Z$+ $4\,b$ SM production. 

Considering these values, we can assess the potential sensitivity of a dedicated analysis using Run 2 LHC data. The typical cross section of the signal is $\sigma(pp\to hhZ )_{ALP}\simeq \left(\frac{c}{f^2_a\textrm{ (TeV)}}\right)^2 \, (230-530) $ fb for all Benchmarks, following the application of basic cuts such as $p_T^b> m_h/2$, $\Delta R_{bb}> 0.4$, and $H_T>$ 500 GeV. Comparing this with the 2 fb cross section from the SM $Z$+ 4$b$ process, and factoring in the b-tagging efficiency of $\epsilon_b=0.77$, we can approximate the significance as follows,\begin{equation}
\frac{S}{\sqrt{B}}\simeq \frac{\sigma_{ALP}\, {\rm Br}(h\to b\bar b)^2 \,\sqrt{{\rm Br}(Z\to 2 \ell)} }{\sqrt{\sigma_{SM}}} \epsilon_b^2\, \sqrt{{\cal L}}  \ ,
\end{equation}
which for the value of the Run 
 2 luminosity ${\cal L}=$140 fb$^{-1}$  and imposing  $S/\sqrt{B}=2$, we would get a limit at 95\% C.L. on the size of ALP-mediated contribution 
\begin{equation}
\label{eq:smoking_limit}
    f_a \gtrsim \textrm{ 2.4-3.0} \times \sqrt{c} \textrm{ TeV} 
\end{equation}
which is  stronger than the limit we had estimated using the di-Higgs channel, see Eq.~\eqref{eq:HMlim}. 

\section{ALP-mediated di-Higgs from top loops}\label{sec:tops}

So far, we have described the ALP coupling to two Higgs bosons and a $Z$ boson based on the chiral expansion discussed in Ref.~\cite{Brivio:2017ije}, which explores a non-linear realization of electroweak symmetry breaking. However, this coupling can also emerge from loop contributions of a linear ALP interacting with the gauge, Higgs, and fermion sectors of the Standard Model, without requiring a non-standard mechanism for electroweak symmetry breaking.

Loop-induced ALP couplings have been discussed in Refs.~\cite{Chala:2020wvs,Bauer:2020jbp,Bonilla:2021ufe,Bonilla:2023dtf,DasBakshi:2023lca}. In particular, in Ref.~\cite{Esser:2023fdo}, loop-induced contributions of the top quark to ALP couplings were investigated, driven by a preference for the ALP to couple with the source of fermion masses. In this context, we are going to consider how the limits from the ALP-mediated contribution to di-Higgs translate into bounds on the coupling of ALPs to top quarks.

\begin{figure}[ht!]
    \centering
    \includegraphics[width=0.65 \textwidth]{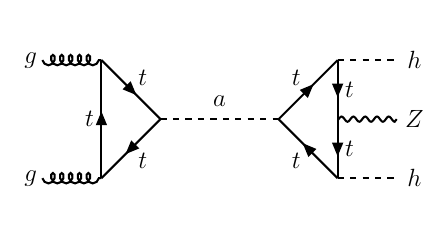}
    \caption{Feynman diagram of the ALP-mediated di-Higgs production through loops of top quarks.}
    \label{fig:top}
\end{figure}

The main diagram contributing to the production of two Higgses in association with a $Z$ boson is shown in Fig.~\ref{fig:top}, where both gluon-fusion and $hhZ$ production depend solely on the ALP-top coupling and the SM couplings of the top quark.

In line with the discussion in Ref.~\cite{Esser:2023fdo}, we will express the ALP-top coupling as follows,
\begin{equation}
\label{eq:top_coupling}
{\cal L} = 
  c_t \,  \frac{\partial_\mu a}{2 f_a} \,  (\bar t \gamma^\mu \gamma^5 t)   \ ,
\end{equation}
where $c_t = c_{t_{R}} - c_{Q_{3}}$ and $t = (t_L, t_R)^T$. It is worth noting that this interaction can also be presented as a mass-dependent coupling,
\begin{equation}
   {\cal L} = 
-i \, c_t \,  \frac{m_t a}{f_a} \,  (\bar t  \gamma^5 t) \ , 
\end{equation}
highlighting that the ALP exhibits a preference for coupling with the top quark over other Standard Model fermions.

Let us first focus on the left part of the diagram in Fig.~\ref{fig:top}. The one-loop contribution of the top to the ALP-gluon coupling has been computed in detail in Ref.~\cite{Bonilla:2021ufe}, and at high momentum transfer is as follows
\begin{equation}\label{ec:estcG}
    c_{\tilde G} \simeq - \frac{\alpha_s}{8 \pi} \frac{c_t}{f_a} \ .
\end{equation}

 On the other hand,  the ALP coupling to two Higgses and a $Z$ boson, the right part of the diagram in Fig.~\ref{fig:top}, has not been computed. A detailed calculation of this loop-induced coupling, including proper renormalization, would require the computation of the diagrams shown in Fig.~\ref{fig:hhZnew}. 
 
\begin{figure}[h!]
    \centering
\begin{tikzpicture}[scale=1]
     \draw[dashed] (0,0)-- (2,1);
     \draw[decoration={aspect=0, segment length=2.5mm, amplitude=0.5mm,coil},decorate] (1,-1) -- (+2,-1);
     \draw[dashed] (-1,0)-- (0,0);
     \draw[->-] (0,0)-- (1,0);
     \draw[->-] (1,0)--(1,-1);
     \draw[->-] (1,-1)--(0,0);
     \draw[dashed] (1,0)-- (2,0);
     \draw (2.25,-1) node {$Z$};
     \draw (2.25,1) node {$h$};
     \draw (2.25,0) node {$h$};
     \draw (-1.25,0) node {$a$};
\end{tikzpicture}\;\;\;\;
\begin{tikzpicture}[scale=1]
     \draw[dashed] (1,1)-- (2,1);
     \draw[decoration={aspect=0, segment length=2.5mm, amplitude=0.5mm,coil},decorate] (1,-1) -- (+2,-1);
     \draw[dashed] (-1,0)-- (0,0);
     \draw[->-] (0,0)-- (1,1);
     \draw[->-] (1,1)--(1,0);
     \draw[->-] (1,0)--(1,-1);
     \draw[->-] (1,-1)--(0,0);
     \draw[dashed] (1,0)-- (2,0);
     \draw (2.25,-1) node {$Z$};
     \draw (2.25,1) node {$h$};
     \draw (2.25,0) node {$h$};
     \draw (-1.25,0) node {$a$};
\end{tikzpicture}
\;\;\;\;
\begin{tikzpicture}[scale=1]
     \draw[dashed] (1,1)-- (2,1);
     \draw[dashed] (1,-1) -- (+2,-1);
     \draw[dashed] (-1,0)-- (0,0);
     \draw[->-] (0,0)-- (1,1);
     \draw[->-] (1,1)--(1,0);
     \draw[->-] (1,0)--(1,-1);
     \draw[->-] (1,-1)--(0,0);
     \draw[decoration={aspect=0, segment length=2.5mm, amplitude=0.5mm,coil},decorate] (1,0)-- (2,0);
     \draw (2.25,-1) node {$h$};
     \draw (2.25,1) node {$h$};
     \draw (2.25,0) node {$Z$};
     \draw (-1.25,0) node {$a$};
\end{tikzpicture}
    \caption{The three diagrams (modulo inverting flow of the top-quark line and permutation of the final state particles) contributing to the one-loop $a$-$h$-$h$-$Z$ production in the linear theory.}
    \label{fig:hhZnew}
\end{figure}
 
The full calculation of this contribution is beyond the scope of this paper, but we will move onto estimating their contribution at high momentum transfer ($\hat s = p_a^2 = m_{hhZ}^2\gg m_{Z,h,t}^2$), which is the kinematic region most interesting for disentangling SM and axion-induced contributions, see the discussion in Sec.~\ref{sec:smoking} and Fig.~\ref{fig:mHHZDRtoleptons}. 

To perform this estimation, we produce the contributing diagrams using FeynArts~\cite{Hahn:2000kx} and the linear ALP model from Ref.~\cite{Brivio:2017ije}.
Before integrating the loop we expand the integrand at order zero in the top, $Z$ and Higgs masses, \textit{i.e.} we perform a Taylor expansion of the integrand in powers of the masses around $m_{Z,h,t}=0$ and take the zeroth order term. Performing the loop integral through standard dimensional regularization formulae in the $\overline{\text{MS}}$ scheme for the resulting integrand, we obtain the (sub)amplitude 
\begin{equation}
    A_{\text{one-top-loop}}^{\mathcal{O}(M^0)}=-\frac{c_{t} y_t^2 g}{(4\pi)^2 f_a c_W}\varepsilon^\ast(p_Z)_\mu\left(\ p_1^\mu \log \frac{(p_1+p_Z)^2}{\tilde{\mu}^2}+\ p_2^\mu \log \frac{(p_2+p_Z)^2}{\tilde{\mu}^2}+\ p_Z^\mu \log \frac{p_Z^2}{\tilde{\mu}^2}\right),\label{eq:chirallooplinear}
\end{equation}
where $\tilde{\mu}=4\pi e^{-\gamma_E} \mu$ is the rescaled renormalisation scale. 

We can now estimate the contribution of this loop to the different Lorentz structures in the Feynman rule, Eq.~\eqref{fig:FR_ahhZ}. For example, for the energy scales of interest we could choose a typical value $\hat{s}=\tilde{\mu}^2=1\;\text{TeV}^2\simeq 16 v^2$. With this choice and further neglecting terms proportional to masses we express the first two terms of Eq.~(\ref{eq:chirallooplinear}) as
\begin{align}
&p_1^\mu\log\frac{(p_1+p_Z)^2}{\hat{s}}=p_1^\mu\log\frac{\hat{s}-2p_2\cdot(p_1+p_Z)-M_Z^2-M_H^2}{\hat{s}}\simeq\nonumber\\&\simeq p_1^\mu \left(\frac{-2p_2\cdot(p_1+p_Z)-M_Z^2-M_H^2}{\hat{s}}\right)\simeq p_1^\mu \left(\frac{-2p_2\cdot(p_1+p_Z)}{\hat{s}}\right)\simeq p_1^\mu \left(\frac{-2p_2\cdot p_a}{16v^2}\right)\\
&p_2^\mu\log\frac{(p_2+p_Z)^2}{\hat{s}}=p_2^\mu\log\frac{\hat{s}-2p_1\cdot(p_2+p_Z)-M_Z^2-M_H^2}{\hat{s}}\simeq\nonumber\\&\simeq p_2^\mu \left(\frac{-2p_1\cdot(p_2+p_Z)-M_Z^2-M_H^2}{\hat{s}}\right)\simeq p_2^\mu \left(\frac{-2p_1\cdot(p_2+p_Z)}{\hat{s}}\right)\simeq p_2^\mu \left(\frac{-2p_1\cdot p_a}{16v^2}\right)
%
%&p_Z^\mu\log\frac{p_Z^2}{\hat{s}}=p_Z^\mu\log\frac{\hat{s}-2p_Z\cdot(p_1+p_2)-p_{hh}^2}{\hat{s}}\simeq\nonumber\\&\simeq 2p_Z^\mu \left(\frac{-p_Z\cdot(p_1+p_2)-p_1\cdot p_2-M_H^2}{\hat{s}}\right)\simeq p_Z^\mu \left(\frac{-2p_1\cdot(p_2+p_Z)}%{\hat{s}}\right)\simeq p_Z^\mu %\left(\frac{-2p_1\cdot(p_a)}{16v^2}\right)
\end{align}
where we expand the logs at NLO in chiral dimensions, $\mathcal{O}(p^2/\hat{s})$.
In this way, adding the two contributions, we get a piece of the $p_{hh}^\mu p_a\cdot p_{hh}$ structure accompanying $\tilde{b}_{14}$ in the chiral ALP EFT $a$-$h$-$h$-$Z$ Feynman rule of Fig.~\ref{fig:FR_ahhZ}. Hence, we assume that the amplitude will contain pieces like
\begin{equation}
    A_{\text{one-top-loop}} \supset \frac{2c_{t} y_t^2 g}{(4\pi)^2 f_a c_W (16v^2)}\varepsilon^\ast(p_Z)_\mu  \, p_{hh}^\mu p_a\cdot p_{hh}.
    %\, \log \hat s \ ,
\end{equation}
Then, the identification of coefficients we estimate is
\begin{equation}\label{ec:est14}
\tilde{b}_{14}\simeq \frac{c_t y_t^2}{16}.
\end{equation}

Placing together these two estimates, Eqs.~\eqref{ec:estcG} and~\eqref{ec:est14}, we can interpret the results from the last section in terms of loop contributions as
\begin{equation}
    \frac{c}{f_a^2}\simeq \frac{ \alpha_s}{128 \pi} \frac{y_t^2 c_t^2}{f_a^2}\ ,
\end{equation}
which needs to be compared with the  limit on $c/f_a^2$ provided  in Eq.~\eqref{eq:HMlim} and the projected Run 2 limit from Eq.~\eqref{eq:smoking_limit}. 

\begin{figure}[t!]
    \centering
    \includegraphics[width=0.9\textwidth]{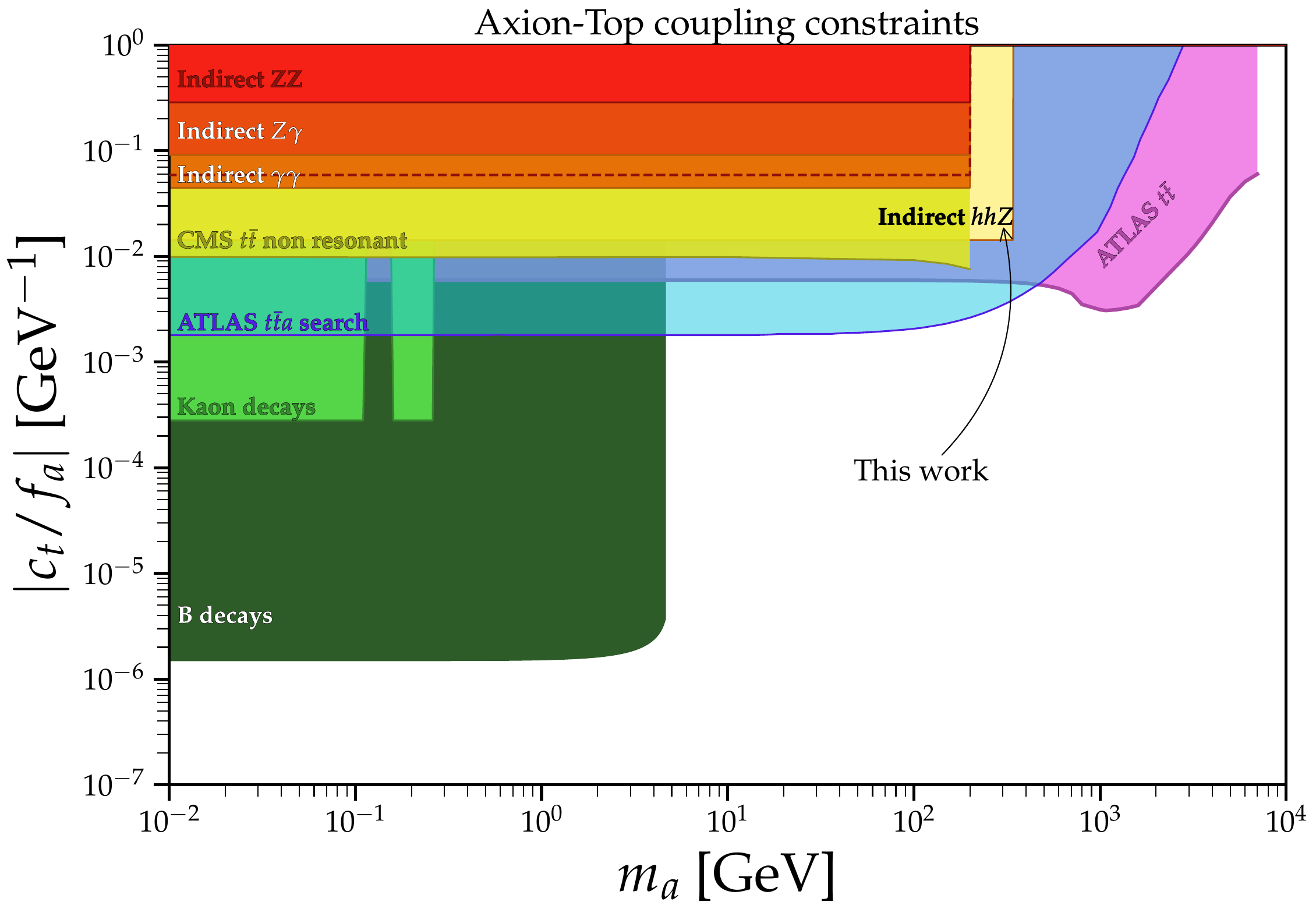}
    \caption{Summary plot of the constraints on ALP-top couplings presented in~\cite{Esser:2023fdo}, to which we have added in cream the indirect constraints on the axion-top coupling deriving from the stronger bound provided in Eq.~\eqref{eq:smoking_limit}. Adapted from Ref.~\cite{Esser:2023fdo} by using the public {\tt AxionLimit} repository~\cite{AxionLimits}. The results of this work can be found at \url{https://cajohare.github.io/AxionLimits/docs/at.html}}
    %namely $f_a\gtrsim\,\sqrt{c}\,3$ TeV
    \label{fig:money2}
\end{figure}

In Fig.~\ref{fig:money2} we show the limit on the ALP-top coupling that the new di-Higgs into $b$-jets search discussed in Sec.~\ref{sec:smoking} could provide, which are stronger than those obtained by the di-Higgs distributions discussed in Sec.~\ref{sec:dihiggs}. We compare the strongest limit derived from Eq.~\eqref{eq:smoking_limit} to the current knowledge from rare decays and LHC's $t\bar t$+missing energy searches for new physics, $t \bar t$ precision measurements and several diboson searches. The reinterpretation of di-Higgs searches is competitive with the indirect searches, even surpassing the  current sensitivity from $t \bar t$ distributions at Run II LHC, and just below the sensitivity of the direct $t \bar t$ plus missing energy search.

\section{Conclusions}
\label{sec:conclusion}

In this paper we have explored the production of a pair of Higgs bosons through a non-resonant ALP.  We confront our theoretical predictions with ATLAS data, establishing competitive constraints on the ALP's couplings to SM particles. This is achieved using a general Effective Field Theory (EFT) framework, employing a chiral Lagrangian that comprehensively describes the interaction between the Electroweak (EW) sector and the ALP candidate.

Due to charge conjugation and parity (CP) conservation of the theory, the production of only two Higgs bosons via an ALP is forbidden (the ALP is a pseudoscalar). This means that the ALP mediated di-Higgs production must be accompanied by additional particles supporting a pseudoscalar coupling: a photon or a $Z$ boson. On-shell, the photon does not contribute to this production mechanism, since it only possesses transverse polarizations. On the other hand, the $Z$ boson does have the longitudinal polarization needed for a CP-allowed on-shell production of $h h Z$ through an ALP. Currently, there are no dedicated searches of two Higgs plus a $Z$ final state, but part of this signal would not be vetoed in the existing di-Higgs searches when the $Z$ boson decays to final-state particles such as hadrons or neutrinos. To show this point, we have reinterpreted an ATLAS experimental analysis~\cite{ATLAS:2023gzn} and provided a rough limit based on the event counting of the {\it high-mass} region. 

We have also argued that one could improve the sensitivity of this the di-Higgs search to ALP-mediated processes by including differential information, in particular of the di-Higgs invariant mass $m_{hh}$, the transverse sphericity, $S_T$ and the $\Delta R_{\gamma \gamma}$ of the two photons produced in the decay of one Higgs boson. Unfortunately, we could not include these distributions as the public information is too limited, but we hope this study would motivate further studies along these lines. Our findings will hopefully encourage experiments to release the information that are needed in order to reinterpret the analyses such as the ones presented in Sec.~\ref{sec:dihiggs}, which employ ML techniques for the signal/background separation and cannot be currently reproduced at all outside the experimental collaborations~\cite{Bierlich:2019rhm,Cranmer:2021urp}.  
Tools like {\tt CONTUR}~\cite{Buckley:2021neu,Butterworth:2016sqg} (which interprets {\tt Rivet}~\cite{Bierlich:2019rhm} analysis outputs) that have access to a broad class of experimental distributions would be instrumental in making more precise statements on the sensitivity of di-Higgs searches to the ALP interactions with the SM bosons, which we motivate in our current work.

When presenting the limits, we have chosen three characteristic benchmarks which should represent the richness of the structure of the ALP coupling to a $Z$ and two Higgs bosons. The first benchmark sets all contributing coefficients to one, the second only turns on the $\tilde{b}_{2D}$ parameter (which is unsuppressed by the chiral counting), and the third only switches on the parameter $\tilde{b}_{17}$ (which is simple enough to test the EFT expansion's validity).

Finally, in the last section we propose a dedicated search of a pair of Higgs bosons and a $Z$ boson as a strong indicator for an ALP mediated production, where the two Higgs bosons decay into two $b\bar{b}$ quark-pairs in association with a leptonic Z. We show various signals for the three benchmarks used in this article. We estimate that one could reach the sensitivity to ALP scales and couplings $f_a/\sqrt{c}$ in the range of 2.4 to 3 TeV with the existing Run 2 LHC luminosity of 140 fb$^{-1}$.

Moreover, the ALP coupling to $hhZ$ can also be understood from a linear ALP coupled to top quarks, leading to loop-induced couplings to gluons and to the $hhZ$ final state. We have estimated these loop contributions and found that the expected limits on the $hhZ$ would place a stringent bound on the ALP-top coupling, competitive with the existing limits~\cite{Esser:2023fdo} from $t\bar t+$ invisible ALP and precision measurements in the $t\bar t$ distributions.

\section*{Aknowledgments}
The research of VS is supported by the Generalitat
Valenciana PROMETEO/2021/083, Proyecto Consolidacion CNS2022-135688, and the Ministerio de Ciencia e
Innovacion PID2020 -113644GB-I00.
F.E. receives funding from the Generalitat Valenciana under the grants GRISOLIAP/2020/145 and PROMETEO/2021/083. ASB acknowledges support from the Generalitat Valenciana under the grant  PROMETEO/2021/083, EU's CNS2022-135688 and PID2022-137003NB-I00 from spanish MCIN/AEI/10.13039/501100011033/  and EU FEDER. ASB gratefully appreciates the hospitality received at IFIC and Universitat de València.  M.U. is supported by the European Research Council under the European Union’s Horizon 2020 research and innovation Programme (grant agreement n.950246), and partially by the STFC consolidated grant ST/T000694/1 and ST/X000664/1.
M.M. is supported by the
Deutsche Forschungsgemeinschaft (DFG, German Research Foundation) under grant 396021762 – TRR 257
Particle Physics Phenomenology after the Higgs Discovery and by the Alexander von Humboldt Foundation.

\newpage
\appendix

\section{ALP-$h$-$h$-$Z$ Feynman rule and kinematics}
\label{app:FR}
In this appendix we will try to simplify the Feynman rule for the $ahhZ$ coupling in Fig.~\ref{fig:FR_ahhZ} by setting all SM particles on-shell and choosing the center of mass (CM) frame. In this frame, the ALP momentum reads $p_a=(\sqrt{s},0,0,0)$ and we choose, by rotational invariance, $p_Z^\mu=(E_Z,0,0,|\boldsymbol{p}_Z|)$ (with $E_Z=\sqrt{m_Z^2+|\boldsymbol{p}_Z|^2}$). Due to conservation of three-momentum, all SM particles' three-momenta must lie in a plane. Hence, the momenta of the two Higgs bosons can be written as $$p_{h1}^\mu=\left(E_{{h1}},|\boldsymbol{p}_{h1}| \sin \varphi,0,|\boldsymbol{p}_{h1}| \cos \varphi\right)$$ and
$$p_{h2}^\mu=\left(\sqrt{s}-E_{{h1}}-E_Z,-|\boldsymbol{p}_{h1}| \sin \varphi,0,-|\boldsymbol{p}_Z|-|\boldsymbol{p}_{h1}| \cos \varphi\right)\;,$$ where $E_{h1}=\sqrt{m_H^2+|\boldsymbol{p}_{h1}|^2}$ and $\varphi$ is the angle between the momenta of the $Z$ boson and the first Higgs boson.

Since we are setting the outgoing particles on-shell, we still need to contract the amplitude with the polarizations of the $Z$ boson, $$\varepsilon^{\pm}_\mu=(0,1,\pm i,0)/\sqrt{2}\;\;\text{ and } \;\;\varepsilon^{L}_\mu=(|\boldsymbol{p}_{Z}|,0,0,E_Z)/m_Z,$$
obtaining
\begin{equation}
  (\varepsilon^+_\mu+\varepsilon^-_\mu)\mathcal{M}^\mu=0
\end{equation}
and (defining $p_{h1}^z\equiv |\boldsymbol{p}_Z|\cos{\varphi}$)
\begin{align}
&\varepsilon^L_\mu\mathcal{M}^\mu\propto e\frac{\sqrt{s}}{m_Z}\Big[-2 {\tilde{a}}_{16}  \left(|\boldsymbol{p}_Z| \left(E_{H_1} \left(E_Z-2 \sqrt{s}\right)+2 E_{H_1}^2\right)-p_{h1}^z \left(E_Z \left(2 E_{H_1}-\sqrt{s}\right)+E_Z^2\right)\right)\nonumber\\
     &+4 {\tilde{a}}_{15}  |\boldsymbol{p}_Z| \left(E_{H_1} \left(\sqrt{s}-E_Z\right)+|\boldsymbol{p}_Z| p_{h1}^z-m_H^2\right)\nonumber+\\
     &+|\boldsymbol{p}_Z| \Big(-16 \pi ^2  v^2 {\tilde{b}}_{2 D}+ {\tilde{b}}_{13}  (m_Z^2+s-2\sqrt{s} E_Z)+{\tilde{b}}_{14}  (s-\sqrt{s} E_Z)+{\tilde{b}}_{12}  \left(s-\sqrt{s} E_Z\right)+{\tilde{b}}_{11}  s\nonumber+\\
     &+{\tilde{b}}_{17}  s+2 \pi s_{2\theta}  m_Z^2  ({\tilde{b}}_3 s_\theta -{\tilde{b}}_{10} c_\theta)/e\Big)\Big].
\end{align}
This simplification is due to CP conservation, and it is discussed in the main text. Also, from this exercise we can see why the $ahh\gamma$ vertex vanishes on-shell: being massless, the photon does not have the contributing longitudinal polarization.

From this treatment we choose the three benchmarks used in this article: first setting all parameters to one, then only $\tilde{b}_{2D}$ being non-zero because it is accompanied by a $(4\pi v)^2$ prefactor and finally only $\tilde{b}_{17}$ since it has the simple behaviour of scaling with $s$.

\section{Effective Operators entering the $ahhZ$ vertex}
\label{app:operators}
The chiral ALP EFT Lagrangian up to NLO contains the operators
\begin{equation}
    {\cal L}_{\rm chiral} \supset c_{2D} \mathcal{O}_{2D} + \sum_{i=\tilde B, \tilde W, \tilde G} \, c_i \, {\cal O}_{i} + \sum_{j=1}^{17} c_j  \, {\cal O}_{j} \, .
\end{equation}
While a full list of all operators can be found in \cite{Brivio:2017ije}, we list here the operators that enter the operators that enter the Feynman rules in Fig.\ \ref{fig:FR_ahhZ}:
\begin{equation}
\begin{split}
    \mathcal{O}_{2D} &= i v^2 \text{Tr} [\mathbf{T V_{\mu}}] \partial^{\mu} \frac{a}{f_a} \mathcal{F}_{2D}(h) \\
    \mathcal{O}_{\tilde{G}} &= - G_{\mu\nu}^a \tilde{G}^{a \mu \nu} \frac{a}{f_a} \\
    \mathcal{O}_3 &= \frac{1}{4\pi} B_{\mu\nu} \partial^{\mu} \frac{a}{f_a} \partial^{\nu} \mathcal{F}_3(h) \\
    \mathcal{O}_{10} &= \frac{1}{4\pi} \text{Tr}[\mathbf{T} W_{\mu\nu}] \partial^{\mu} \frac{a}{f_a} \partial^{\nu} \mathcal{F}_{10}(h)\\
    \mathcal{O}_{11} &= \frac{i}{(4\pi)^2} \text{Tr} [\mathbf{T V_{\mu}}] \Box \frac{a}{f_a} \partial^{\mu} \mathcal{F}_{11}(h) \\
    \mathcal{O}_{12} &= \frac{i}{(4\pi)^2} \text{Tr} [\mathbf{T V_{\mu}}] \partial^{\mu} \partial^{\nu} \frac{a}{f_a} \partial_{\nu} \mathcal{F}_{12}(h) \\
    \mathcal{O}_{13} &= \frac{i}{(4\pi)^2} \text{Tr} [\mathbf{T V_{\mu}}]  \partial^{\mu}\frac{a}{f_a} \Box \mathcal{F}_{13}(h) \\
     \mathcal{O}_{14} &= \frac{i}{(4\pi)^2} \text{Tr} [\mathbf{T V_{\mu}}]  \partial_{\nu} \frac{a}{f_a} \partial^{\mu} \partial^{\nu} \mathcal{F}_{14}(h) \\
      \mathcal{O}_{15} &= \frac{i}{(4\pi)^2} \text{Tr} [\mathbf{T V_{\mu}}]  \partial^{\mu}\frac{a}{f_a} \partial_{\nu} \mathcal{F}_{15}(h) \partial^{\nu} \mathcal{F}_{15}^{'}(h)\\
      \mathcal{O}_{16} &= \frac{i}{(4\pi)^2} \text{Tr} [\mathbf{T V_{\mu}}]  \partial_{\nu}\frac{a}{f_a} \partial^{\mu} \mathcal{F}_{16}(h) \partial^{\nu} \mathcal{F}_{16}^{'}(h)\\
      \mathcal{O}_{17} &= \frac{i}{(4\pi)^2} \text{Tr} [\mathbf{T V_{\mu}}] \partial^{\mu} \Box \frac{a}{f_a} \mathcal{F}_{17}(h). \\
\end{split}    
\end{equation}
$a_i$ and $b_i$ denote the linear and quadratic order in $h/v$ of the Higgs flare function $\mathcal{F}_i$, cf.\ Eq.~(\ref{ec:poly}). For convenience, we write $\tilde{a}_i = c_i a_i$ and $\tilde{b}_i = c_i b_i$.

\bibliographystyle{JHEP}
\bibliography{references.bib}
\end{document}